\documentclass[10pt,conference]{IEEEtran}
\usepackage{epsfig,rotating,setspace,amsmath,epsf,amssymb,amsfonts,bm,bbm,theorem,cite,graphicx,epstopdf,algorithm,algpseudocode,float,color,mathtools,authblk,tikz,physics,nicefrac,xcolor,comment,subfigure}

\IEEEoverridecommandlockouts
\allowdisplaybreaks

\begin{document}

\title{How to Maximize Efficiency in Systems with Exhausted Workers
\thanks{Research of MB was supported in part by Tubitak Bilgem EDGE-4-IoT and Tubitak 2232-B program.}}

\author[1]{Elif Beray Sariisik}
\author[1]{Melih Bastopcu}
\author[1]{Nail Akar}
\author[2]{Sennur Ulukus}

\affil[1]{\normalsize Department of Electrical and Electronics Engineering, Bilkent University, Ankara, T\"{u}rkiye}
\affil[2]{\normalsize Department of Electrical and Computer Engineering, University of Maryland, College Park, MD, USA}

\maketitle

\begin{abstract}
  We consider the problem of assigning tasks efficiently to a set of workers that can exhaust themselves as a result of processing tasks. If a worker is exhausted, it will take a longer time to recover. To model efficiency of workers with exhaustion, we use a continuous-time Markov chain (CTMC). By taking samples from the internal states of the workers, the source assigns tasks to the workers when they are found to be in their efficient states. We consider two different settings where (i) the source can assign tasks to the workers only when they are in their most efficient state, and (ii) it can assign tasks to workers when they are also moderately efficient in spite of a potentially reduced success probability. In the former case, we find the optimal policy to be a threshold-based sampling policy where the thresholds depend on the workers' recovery and exhaustion rates. In the latter case, we solve a non-convex sum-of-ratios problem using a branch-and-bound approach which performs well compared with the globally optimal solution.      
\end{abstract}

\section{Introduction}
Timely and informed decision-making is fundamental in dynamic distributed systems, particularly when the efficiency of task execution depends on the evolving internal conditions of system components. In real-world settings, worker nodes often exhibit state-dependent behavior influenced by fatigue, recovery, and other intrinsic stochastic processes. These concepts have been implemented in a range of systems, including state-aware agent systems \cite{Bochkovskyi2024workerfatigue}, degradation-aware machine models \cite{Kumar2023, cholette2014machine}, control and communication systems \cite{scheuvens2021control, channels_that_die}, cloud platforms \cite{zhou2022server}, and energy-harvesting systems \cite{chen2014energyharvest}. Without accurate tracking of these evolving internal states, task allocation policies may fail to capture the true operational capacity of the workers. Age of information (AoI) has been widely employed to quantify the timeliness of systems \cite{Yates2021JSAC}. However, in many practical settings, focusing only on reducing AoI may be insufficient where the effectiveness of an update is tightly linked to the evolving internal state of the system.

\begin{figure}[t]
    \centering
    \includegraphics[width=0.7\columnwidth]{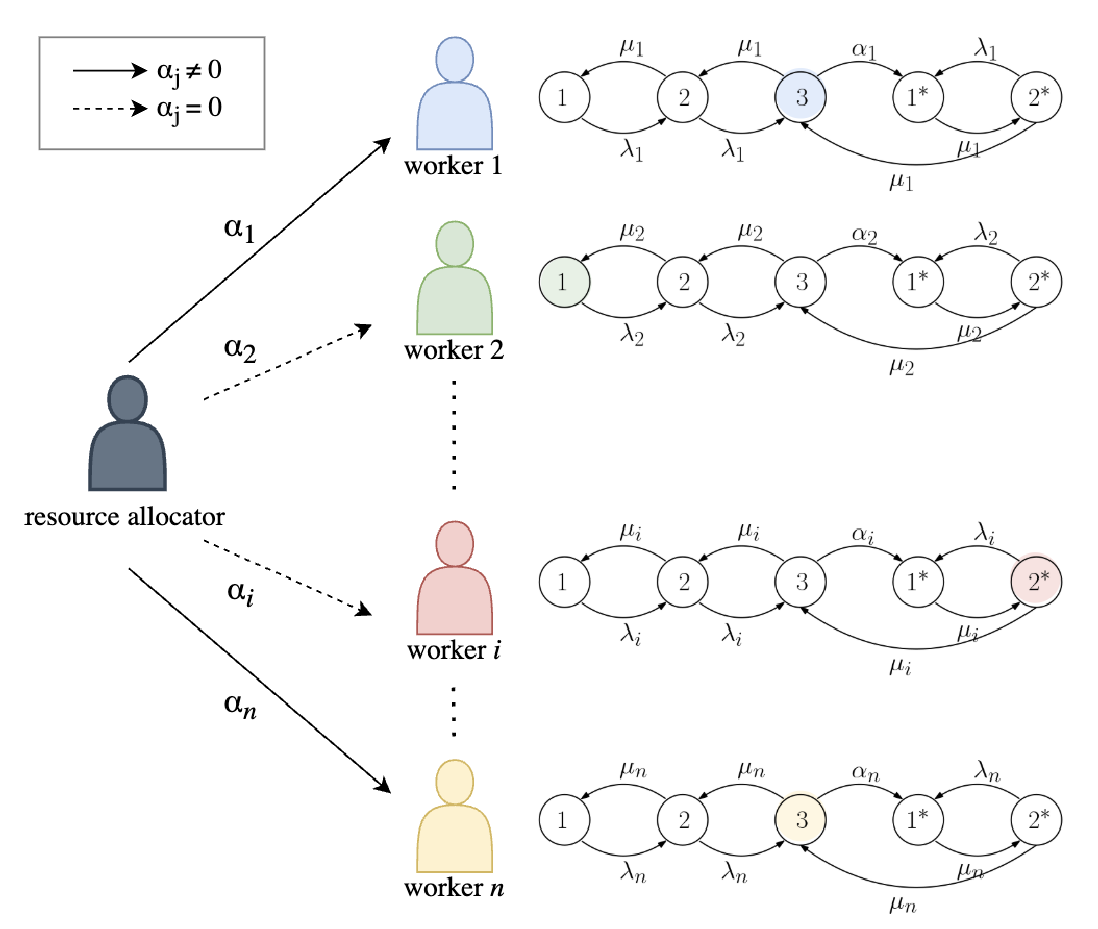}
    \vspace*{-0.2cm}
    \caption{A centralized resource allocator (source) assigning tasks to \( n \) workers with rates $\alpha_i$'s, only when each worker is found to be efficient. Each worker follows a Markov process with transitions among external states \( 1 \), \( 2 \), and \( 3 \), and internal processing states \( 1^* \) and \( 2^* \), driven by recovery rate (\( \lambda \)) and exhaustion rate (\( \mu \)). Shaded states indicate current worker states.}
    \label{fig:sytem_model}
    \vspace{-0.5cm}
\end{figure}

Motivated by this, in this work, we consider a system where the source assigns timely tasks to $n$ workers that can exhaust themselves as a result of processing tasks as illustrated in Fig.~\ref{fig:sytem_model}. We model exhaustion and recovery of the workers with a continuous-time Markov chain (CTMC) where each worker has \textit{inefficient, moderately efficient,} and \textit{highly efficient} states to process tasks, denoted by states 1, 2, and 3, respectively. If workers process a task, then their internal states move to special exhaustion states denoted by $1^*$, where it will take a longer time duration to recover. Knowing this mechanism, the source wants to track the internal states of the workers and assign tasks to them whenever they are efficient.        

Our work is closely related to the timely remote estimation of CTMC which has been considered in the literature \cite{Maatouk_remote_est, Akar2024_CTMC, Pappas2024c, Pappas2024b, Pappas2024a}. In particular, \cite{Maatouk_remote_est} introduces the age of incorrect information (AoII) metric where the underlying information is dynamically modeled as a CTMC, \cite{Akar2024_CTMC} considers the remote estimation of multiple heterogeneous CTMCs with a binary freshness metric, and \cite{Pappas2024c, Pappas2024b, Pappas2024a} focus on optimum sampling policies to timely track CTMCs under different resource constraints. In all these works, the underlying Markov chain evolves independently from the sampling process. However, in our work, since the source assigns tasks to workers resulting in exhaustion, the sampling policy also affects the steady-state distribution of the workers. 

The works that are most closely related to our work are \cite{liyanaarachchi2025mm} and \cite{banerjee2025mm}. Reference \cite{liyanaarachchi2025mm} considers the problem of external job assignments to multiple Markov machines (MMs) where each machine moves in between idle and busy states. By using the binary freshness metric, the goal of the source is to track the internal states of MMs to efficiently assign tasks to the MMs under a total sampling constraint. Reference \cite{banerjee2025mm} considers a similar framework but focuses on a single MM, where the MM is assigned tasks by a server which is penalized if the assigned tasks are dropped due to the busy state of the MM. \cite{banerjee2025mm} uses a Markov decision process (MDP) framework to minimize the AoII and also the penalty due to dropped tasks. 

In this work, we propose a sampling rate allocation strategy for systems in which the source has limited access to the internal states of the workers due to a global sampling budget. Each worker follows a stochastic five-state model $\{1,2,3,1^*,2^*\}$, where the state represents the worker’s ability to execute tasks. 
Tasks can only be successfully completed when the worker is in certain states, and assigning a task causes task-induced exhaustion, which triggers a recovery process modeled as a CTMC. We consider two scenarios: (i) the source assigns tasks only when the worker is fully efficient, and (ii) the source can assign tasks also when the worker is moderately efficient, but with a potentially reduced success probability. We derive the optimum (the sub-optimum) sampling rate allocations for the former (for the latter) case that maximize the overall task success rate under a total sampling budget constraint. In the first case, the resulting optimization is convex and admits a threshold-based structure; in the second case, we solve a non-convex sum-of-ratios problem using a branch-and-bound approach. Our results show that the optimal sampling policies prioritize workers based on their exhaustion-to-recovery dynamics and task success probabilities, and in some regimes, it may be optimal to completely ignore certain workers.

\section{System Model and Problem Formulation}
In this work, we consider a system that consists of a source and $n$ worker nodes. We model each worker to have multiple levels of \textit{efficiency} to \textit{successfully complete} an assigned task from the source. When there is no task assigned, the workers' levels of efficiency change over time ranging from the highly efficient state where it completes the task successfully with probability 1 and to a completely inefficient state where it is unable to perform any task. Thus, when there is a task assigned, workers will compete the tasks based on their current level of efficiencies. As a consequence, they may exhaust themselves, transitioning into special \textit{exhaustion states}, where recovery to normal efficiency levels takes significantly longer. 

To capture the varying efficiency levels of workers, we model them using a representative 5-state CTMC with states $S = \{1,2,3,1^*,2^*\}$.\footnote{A more refined CTMC with additional efficiency levels could be considered, which we leave as a direction for future research.} Here, the states 1, 2, and 3 represent the \textit{inefficient,} \textit{moderately efficient}, and \textit{highly efficient} levels, respectively, while states $1^*$ and $2^*$ denote the \textit{inefficient,} \textit{moderately efficient} levels when the workers are in an exhausted state. We denote worker $i$'s level of efficiency at time $t$ as $s_i(t) \in S$ for $i\in\{1,\dots, n\}$. Worker $i$ transitions from state $k-1$ to state $k$ (for $k = 2,3$) at a rate of $\lambda_i$, referred to as the worker's \textit{recovery rate}. Similarly, worker $i$ transitions from state $k$ to state $k-1$ (for $k = 2,3$) at a rate of $ \mu_i$, referred to as the worker's \textit{exhaustion rate}. We assume $ \lambda_i \geq \mu_i$, indicating that the recovery rate exceeds the exhaustion rate when no tasks are assigned. As a result of processing tasks, the workers are exhausted and they transition to state $1^*$. During recovery, worker $i$ transitions from state $1^*$ to $2^*$ and eventually from state $2^*$ to $3$ (fully recovered) at a rate of $\mu_i$ and transitions from state state $2^*$ to state $1^*$ at a rate of $\lambda_i$. As we assume that $\lambda_i\geq \mu_i$, worker $i$ recovers at a slower rate after processing a task compared to its normal state evolution. 

Worker $i$ can process the tasks when $s_i(t)\in \{2,2^*,3\}$. If the internal state of the worker node is equal to $s_i(t)=3$, then the tasks can be completed perfectly without an error. If the state is equal to $s_i(t)=\{2,2^*\}$, that is, when the worker is moderately efficient, tasks can be successfully completed with probability $p_{s,i}$ where $0\leq p_{s,i}\leq 1$. The source wants to assign tasks to the worker nodes and successfully process them. For that, the source takes samples from the internal states of the workers.\footnote{In our model, we consider a setting where the source \textit{randomly} probes the workers' states and assigns tasks to them when they are efficient. In the literature, there are settings where the source can \textit{continuously} probe the state evolutions and decide an action for each state as in \cite{Tang_MDP}. Another viable option is randomly assigning tasks to workers without any probing with rate $\alpha_i$.} We model the source's sampling as a Poisson process with rate $\alpha_i\geq 0$. In an ideal scenario, the source should sample each worker's state all the time to assign tasks as much as possible. However, due to resource constraints, the total sampling rate of the source is limited and we have $\sum_{i=1}^{n} \alpha_i \leq C$. Due to the worker's state evolution and the sampling process, the states $s_i(t)\in \mathcal{S}$ form a CTMC where the stationary distribution is given by $\pi_{j}$ where $j\in \mathcal{S}$. The goal of the source is to \textit{successfully} process as many tasks as possible at the worker nodes. 

Next, we will consider the case where the source assigns tasks to workers only when they are highly efficient. 

\begin{figure}[t]
    \centering
    \includegraphics[width=0.65\columnwidth]{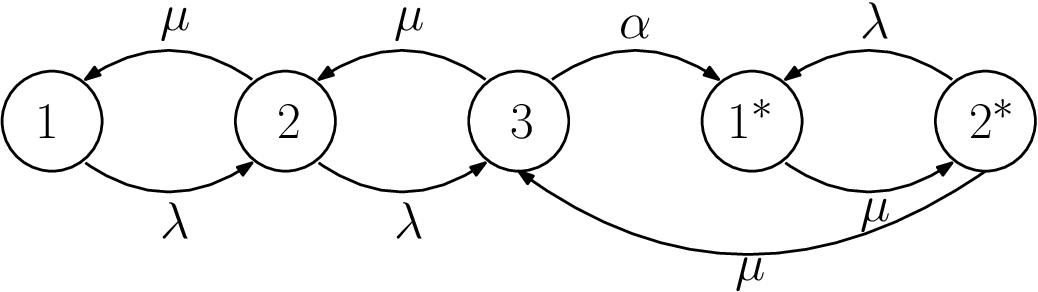}
    \vspace{-0.2cm}
    \caption{State transitions in the CTMC model of a worker. Here, we present a general worker's states as $s(t)$ and their transition rates.}
    \label{fig:CTMC_1}
    \vspace{-0.5cm}
\end{figure}

\section{Task Processing Only When the Workers are Most Efficient} \label{sect:most}
First, we focus our attention to the case where the tasks are assigned to the workers only when they are most efficient, i.e., $s_i(t) = 3$. In this case, we present the CTMC's states and their transition rates for a generic worker (where we omit the worker $i$'s index for simplicity) in Fig.~\ref{fig:CTMC_1}. In order to find the stationary distribution of the CTMC shown in Fig.~\ref{fig:CTMC_1}, we write the state balance equation as follows:
\begin{align}
    \lambda \pi_{1} &= \mu \pi_{2}, \label{eqn:state-balance-1} \\
    (\lambda +\mu) \pi_{2} &= \lambda \pi_{1}  + \mu \pi_{3}, \\
    (\mu +\alpha) \pi_{3} &= \lambda \pi_{2} +\mu \pi_{2^*} , \\
    \mu  \pi_{1^*} &= \lambda \pi_{2^*} + \alpha \pi_{3}, \\
    (\lambda + \mu)  \pi_{2^*} &= \mu \pi_{1^*} .\label{eqn:state-balance-10}
\end{align}
Then, solving (\ref{eqn:state-balance-1})-(\ref{eqn:state-balance-10}) jointly with $\sum_{j\in \mathcal{S}} \pi_{j} = 1$ gives us the steady state distribution of $\pi_{j}$'s as $\pi_1 = \frac{\mu^4}{K},$ $\pi_2 = \frac{\lambda \mu^3}{K},$ $\pi_3 = \frac{\lambda^2 \mu^2}{K},$ $\pi_{1^*} = \frac{\alpha \lambda^2 (\lambda +\mu)}{K},$ $\pi_{2^*} = \frac{\alpha \lambda^2\mu}{K}$ where $K=\alpha \lambda^3 + \lambda^2 \mu^2 + 2\alpha \lambda^2 \mu + \lambda \mu^3 +\mu^4$. After obtaining the steady-state distribution, we note that each worker processes tasks while moving from state $3$ to state $1^*$, that is, $\alpha \pi_{3}$. Thus, by placing workers' index back, we formulate the source's task assignment problem as: 
\begin{align}
    \max_{\{ \alpha_i \}}  \quad & \sum_{i=1}^{n} \frac{\lambda_i^2\mu_i^2}{\lambda_i^3+2\lambda_i^2\mu_i +\frac{\lambda_i^2\mu_i^2 +\lambda_i \mu_i^3+\mu_i^4}{\alpha_i}} \nonumber \\[-3pt]
    \mbox{s.t.} \quad & \sum_{i=1}^{n} \alpha_i \leq C \nonumber \\[-2.5pt]
    \quad & \alpha_i\geq 0, \quad i\in\{1,\ldots,n \}. \label{problem1}
\end{align}
The objective function in (\ref{problem1}) is equal to $\alpha \pi_{3}$ and $\sum_{i=1}^{n} \alpha_i \leq C$ is the total sampling rate constraint. As the objective function in (\ref{problem1}) is a concave function of $\alpha_i$ and the constraints are convex, the optimization problem in (\ref{problem1}) is convex \cite{Boyd04}. Thus, we define the Lagrangian function as
\begin{align}
    \mathcal{L} = -\sum_{i=1}^{n}\frac{\lambda_i^2\mu_i^2}{B_i + \frac{A_i}{\alpha_i}} +\beta \left(\sum_{i=1}^{n} \alpha_i - C \right) -\sum_{i=1}^{n} \nu_i\alpha_i,  
\end{align}
where $\beta\geq 0$ and $\nu_i\geq 0$ for $i = 1,\dots, n$ are the Lagrange multipliers and we defined $A_i = \lambda_i^2\mu_i^2 +\lambda_i \mu_i^3+\mu_i^4$ and $B_i = \lambda_i^3+2\lambda_i^2\mu_i $. 
Next, the KKT conditions are given by
\begin{align}\label{eqn:KKT}
    \frac{\partial\mathcal{L}}{\partial \alpha_i} =& -\frac{\lambda_i^2\mu_i^2 A_i}{(A_i + \alpha_i B_i)^2} + \beta - \nu_i = 0,
\end{align}
and complementary slackness (CS) conditions are given by
\begin{align}
\beta\left( \sum_{i=1}^{n} \alpha_i - C \right) =& 0, \label{eqn:CS1} \\ \nu_i \alpha_i  =& 0, \quad i = 1,\dots, n.\label{eqn:CS2}
\end{align}
By using (\ref{eqn:KKT}), we obtain $\alpha_i$ as
\begin{align}\label{eqn:alpha_i_int_soln}
    \alpha_i =
    \frac{A_i}{B_i}\left(\frac{\lambda_i\mu_i }{\sqrt{A_i(\beta - \nu_i)}}-1\right).
\end{align}
By using the CS conditions in (\ref{eqn:CS2}), we have either $\alpha_i>0$ and $\nu_i = 0$ or $\alpha_i=0$ and $\nu_i \geq 0$. Thus, we rewrite (\ref{eqn:alpha_i_int_soln}) as
\begin{align}\label{eqn:alpha_i_final}
    \!\!\alpha_i \!\!=\!\!
    \frac{A_i}{B_i}\left(\frac{\lambda_i\mu_i }{\sqrt{A_i\beta }}-1\right)^+ \!\!\!\!\!=\!  \frac{A_i}{B_i} \left(\frac{1}{\sqrt{\beta}}\frac{1}{\sqrt{1\!+\!\frac{\mu_i}{\lambda_i} \!+\!\frac{\mu_i^2}{\lambda_i^2}}}-1\right)^+\!\!\!,\!\!
\end{align}
where $(\cdot)^+ = \max\{\cdot, 0\}.$ Finally, we can find  $\beta$ that will honor the total sampling budget constraint, $\sum_{i}^{n}\alpha_i =C$ by following similar solution steps as provided in \cite{bastopcu_binary_freshness_21}. In summary, to find $\alpha_i$ values, we will first neglect $(\cdot)^+$ operator in (\ref{eqn:alpha_i_final}), and define  $\hat{\alpha}_i =  \frac{A_i}{B_i}\left( \frac{1}{\sqrt{\beta}}\left(\sqrt{1\!+\!\frac{\mu_i}{\lambda_i} \!+\!\frac{\mu_i^2}{\lambda_i^2}}\right)^{-1}-1\right)$. Then, we obtain $\beta$ by solving $\sum_{i}^{n}\hat{\alpha}_i =C$. We will substitute $\beta$ inside $\hat{\alpha}_i$ and check whether $\hat{\alpha}_i\geq 0$ for all $i$. If all $\hat{\alpha}_i$'s are nonnegative, then we find the optimal $\alpha_i$ allocations via $\alpha_i = \hat{\alpha}_i$ for all $i$. If some of the $\hat{\alpha}_i$ terms are negative, then we choose the worker with the highest $1+\frac{\mu_i}{\lambda_i} +\frac{\mu_i^2}{\lambda_i^2}$ terms and set $\hat{\alpha}_i =0$. Then, we will solve $\sum_{i}^{n}\hat{\alpha}_i =C$ for $\beta$ with the remaining $\hat{\alpha}_i$ terms. We will continue to repeat this process until all $\hat{\alpha}_i$ terms are non-negative. When this condition is satisfied, we obtain the final solution by setting $\alpha_i = \hat{\alpha}_i$ for all $i$. Here, we note that the optimal sampling rate allocation policy is a threshold policy, that is, the source may not assign any tasks to some of the workers. These workers could be the ones that have the highest $1+\frac{\mu_i}{\lambda_i} +\frac{\mu_i^2}{\lambda_i^2}$ terms which is equivalent to the workers with the highest $\frac{\mu_i}{\lambda_i}$ values. That is, if the workers have relatively high exhaustion rate $\mu_i$ compared to their own recovery rate $\lambda_i$, then the source may not utilize these workers since it may take longer time for recovery especially when the source's total sampling rate $C$ is limited. Finally, once the worker has a positive sampling rate $\alpha_i$, their overall optimal rate allocations are scaled with $\frac{A_i}{B_i}$ terms in (\ref{eqn:alpha_i_final}) which is different than their relative exhaustion-recovery ratio $\frac{\mu_i}{\lambda_i}$. 

In the next section, we consider a scenario, where the source can assign tasks to the worker nodes when their internal state is equal to $s_i(t) = 2$ or $s_i(t) = 2^*$. 

\begin{figure}[t]
    \centering
    \includegraphics[width=0.7\columnwidth]{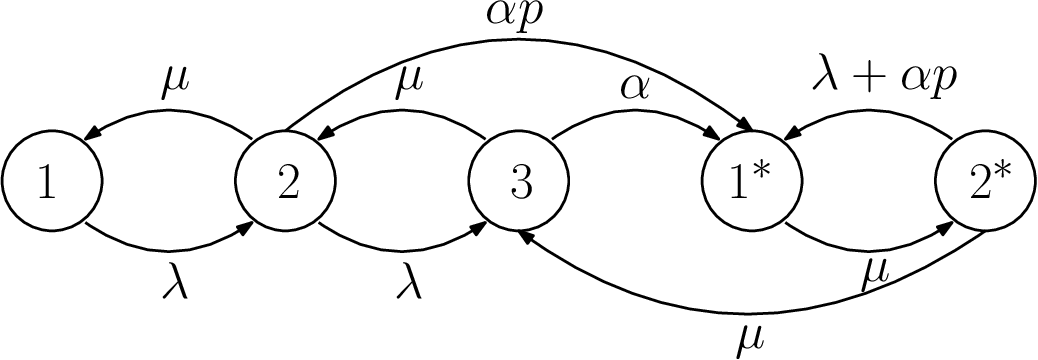}
    \vspace{-0.2cm}
    \caption{State transitions in the CTMC model of a worker when the source assigns tasks with probability $p$ if a worker is moderately efficient. Here, we present a general worker's states as $s(t)$ and their transition rates.}
    \label{fig:CTMC_2}
    \vspace{-0.5cm}
\end{figure}

\section{Tasks Processing When the Workers are Moderately Efficient} \label{sect:moderate}
In this section, we consider the scenario where the source can assign tasks to the workers even when the workers are moderately efficient, that is, the source can assign tasks to the workers when their states are in $s_i(t) \in \{2,3,2^*\}$. Different from completing a task when the worker is most efficient, i.e., $s_i(t) =3$, worker $i$ successfully completes tasks with probability $p_{s,i}$ where $0\leq p_{s,i}\leq 1$ when they are moderately efficient, i.e., when $s_i(t)\in \{2, 2^*\}$. Thus, especially when $p_{s,i}$ is low, the source may not be willing to assign tasks to a worker when it is moderately efficient. As a result, the source assigns tasks with probability $p_i$ where $0\leq p_i\leq 1$ when the source observes that $s_i(t) \in \{2,2^*\}$. Then, the CTMC describing the state-space and their corresponding transitions are shown in Fig.~\ref{fig:CTMC_2}. Different from the CTMC discussed in the previous section shown in Fig.~\ref{fig:CTMC_1}, we see the additional transition rates $\alpha p$ going from states 2 and $2^*$ to the state $1^*$, representing the task assignments at the moderate efficiency.  

Similar to the previous section, we first find the stationary distribution of the CTMC (denoted by $\!\Tilde{\pi}_j$, $j\!\in \mathcal{S}$) shown in Fig.~\ref{fig:CTMC_2}. For that, the state-balance equations are given as
\begin{align}
 \lambda \Tilde{\pi}_1 &= \mu \Tilde{\pi}_2,\label{eqn:state-balance-1-tilde} \\   
 (\mu+\lambda + \alpha p )\Tilde{\pi}_2 &=  \lambda \Tilde{\pi}_1 + \mu \Tilde{\pi}_3,  \\
 (\mu+\alpha) \Tilde{\pi}_3 &= \lambda \Tilde{\pi}_2 + \mu \Tilde{\pi}_{2^*}, \\
 \mu \Tilde{\pi}_{1^*} &= \alpha \Tilde{\pi}_3 + \alpha p \Tilde{\pi}_2 + (\lambda +\alpha p) \Tilde{\pi}_{2^*},\\
 (\mu+\lambda +\alpha p) \Tilde{\pi}_{2^*} &= \mu \Tilde{\pi}_{1^*}.\label{eqn:state-balance-last-tilde}
\end{align}

By using (\ref{eqn:state-balance-1-tilde})-(\ref{eqn:state-balance-last-tilde}) and the fact that $\sum_{j\in\mathcal{S} } \Tilde{\pi}_j = 1$, we can find the steady-state distribution $\Tilde{\pi}_j$. Due to the complexity of the expressions, we do not provide the closed-form formulas for $\Tilde{\pi}_j$'s here. After finding the steady-state distribution, we can write the source's objective function which is to maximize the overall task completion efficiency given by
\begin{align}\label{eqn:obj_fnc2}
  \alpha \Tilde{\pi}_{3} + p_s \alpha p (\Tilde{\pi}_{2^*}+\Tilde{\pi}_{2}).  
\end{align}
After substituting the closed form expressions for steady-state probabilities in (\ref{eqn:obj_fnc2}) and placing the worker index $i$ back, we can obtain the optimization problem given in (\ref{problem2}). 
\begin{figure*}[t]
    \begin{align}
    \label{problem2}
    \max_{\{ \alpha_i, p_i \}}  \quad & \sum_{i=1}^{n}\frac{f_i(\alpha)}{g_i(\alpha)} = \sum_{i=1}^{n} \frac{\alpha_i^3 \frac{p_{s,i} p_i^2 \lambda_i}{\mu_i^3}+\alpha_i^2 \frac{p_i \lambda_i}{\mu_i^2} \left(1+p_i p_{s,i} + \frac{p_{s,i}\lambda_i}{\mu_i}\right)+\alpha_i \left(\frac{p_i\lambda_i p_{s,i}}{\mu_i} + \frac{\lambda_i^2}{\mu^2_i}\right) }{\alpha_i^3\frac{\lambda_i p_i^2}{\mu_i^4}+\alpha_i^2 \frac{\lambda_i p_i}{\mu_i^3}\left(2+\frac{2\lambda_i}{\mu_i} +p_i\right)+\alpha_i \frac{\lambda_i}{\mu_i^2}\left(\frac{\lambda_i^2}{\mu_i^2} +\frac{\lambda_i}{\mu_i}(2+p_i)+2p_i\right)+1+\frac{\lambda_i}{\mu_i}+\frac{\lambda_i^2}{\mu_i^2}} \nonumber \\
    \mbox{s.t.} \quad & \sum_{i=1}^{n} \alpha_i \leq C \nonumber \\
    \quad & \alpha_i\geq 0, \quad i\in\{1,\ldots,n \}\nonumber \\
    \quad & 0\leq p_i\leq 1, \quad i\in\{1,\ldots,n \}.
    \end{align}
    \hrule 
\end{figure*}

We note that the goal is to maximize the task efficiency of the workers via optimizing over the sampling rate $\alpha_i$ and the task assignment probability at the moderate efficiency $p_i$ subject to the total sampling rate constraint $ \sum_{i=1}^{n} \alpha_i \leq C$ and some feasibility constraints. The objective function in (\ref{problem2}) is non-convex in terms of the optimization variables $\alpha_i$'s and $p_i$'s. To solve (\ref{problem2}), for a given set of $p_i$'s, we will provide a solution for $\alpha_i$ and then search iteratively over the optimal $p_i$'s. The optimization problem in (\ref{problem2}) falls into a well-known \textit{sum-of-ratios problem} in the literature \cite{jong2012practical, khorramabadi2021sum}. Furthermore, for a given set of $p_i$'s, both the numerator and denominator are convex functions of $\alpha_i$'s over the feasible region $ \sum_{i=1}^{n} \alpha_i \leq C$ and $\alpha_i\geq 0$. Similarly, for a given set of $\alpha_i$'s, both the numerator and denominator are convex functions of $p_i$'s over the feasible region $ 0\leq p_i \leq 1$ for all $i$. Thus, in each iteration, we will focus on solution methods provided in \cite{SHEN20132301,SHEN2009145}. 

To solve the optimization problem in (\ref{problem2}), we employ a branch-and-bound algorithm, which iteratively refines the search space while maintaining upper and lower bounds on the objective function. The algorithm operates by partitioning the feasible region into progressively smaller simplexes, thereby narrowing down the optimal values of the optimization variables. The detailed steps of the algorithm are provided in Algorithm~\ref{Alg1}, which is based on methods given in \cite{SHEN20132301, SHEN2009145}. 

\begin{algorithm}[t]
    \caption{Branch-and-Bound Algorithm for Sum-of-Ratios Optimization in \cite{SHEN20132301}}\label{Alg1}
    \begin{algorithmic}[1]
    
    \State \textbf{Initialization:}
    \State $S_0\! \!= \!\!\{V_0, V_1, \dots, V_n\},~ LB_0\! \gets\! -\infty, ~ \!k\! \gets \!0, ~ Q_0 \!=\! \{S_0\}$
    \State $\rho, C, \lambda_i, \mu_i, p_{s,i}$ given
    \State $\hat{s}^{S_0}\!\gets\!\frac{1}{n+1}\sum_{i=0}^n V_i^{S_0}$
    \State $M_j^{S_0}\!\gets\!\max\{g_j(V_i^{S_0}):i=0,\dots,n\}$
    \State $m_j \gets \min \{ g_j(x) : x \in X \}$ 
    
    \State \textbf{Solve $P_3(S_0)$:}
    \State \qquad $UB_0\!=\!\max\sum_{j=1}^m\sum_{i=0}^n\beta_{ji}^{S_0}f_j(V_i^{S_0})$
    \State \qquad \text{s.t.} $\frac{1}{M_j^{S_0}}\leq u_j^{S_0}\leq\frac{1}{m_j},\;\forall j$
    \State \qquad $\sum_{i=0}^n p_j^s V_i^{S_0}\beta_{ji}^{S_0}{+}(g_j(\hat{s}^{S_0}){-}p_j^s\hat{s}^{S_0})u_j^{S_0}\leq1,\;\forall j$
    \State \qquad $\sum_{i=0}^n\beta_{ji}^{S_0}-u_j^{S_0}=0,\;\forall j$
    \State \qquad $\sum_{i=0}^n d_q^s V_i^{S_0}\beta_{ji}^{S_0}{+}(\phi_q^s{-}d_q^s\hat{s}^{S_0})u_j^{S_0}\leq0,\;\forall j,q$
    \State \qquad $\beta_{ji}^{S_0}\geq0,\;\forall j,i$
     
    \State $x^{S_0} = \sum_{i=0}^{n} \beta_{ji}^{S_0} V_i^{S_0} / u_j^{S_0}$    
    \State $LB_0 \gets \sum_{j=1}^m \frac{f_j(x^{S_0})}{g_j(x^{S_0})}$
    
    \State \textbf{While} $UB_k - LB_{k-1} > \rho UB_k$
    \State \qquad $S^*\!\gets\!\arg\max_{S\in Q_k}\text{edge}(S)$
    \State \qquad $\hat{s}^{S^*}\!\gets\!\frac{1}{n+1}\sum_{i=0}^n V_i^{S^*}$
    \State \qquad $M_j^{S^*}\!\gets\!\max\{g_j(V_i^{S^*}):i=0,\dots,n\}$
    \State \qquad $m_j\!\gets\!\min\{g_j(x):x\in X\}$
    
    \State \qquad \textbf{Solve $P_3(S^*)$:}
    \State \qquad\quad $UB^*\!=\!\max\sum_{j=1}^m\sum_{i=0}^n\beta_{ji}^{S^*}f_j(V_i^{S^*})$
    \State \qquad\quad \text{s.t. constraints in $P_3(S_0)$ applied to $S^*$}
    
    \State \qquad $x^{S^*}\!=\!\sum_{i=0}^{n}\beta_{ji}^{S^*}V_i^{S^*}/u_j^{S^*}$
    \State \qquad
    $LB^*\!\gets\!\sum_{j=1}^m\!f_j(x^{S^*})/g_j(x^{S^*})$
    
    \If{$LB^* > LB_k$}
        \State \qquad$LB_k \gets LB^*$
        \State \qquad $x^* \gets x^{S^*}$
    
    \EndIf
    \State \qquad $c \gets$ midpoint$(S^*)$
    \State \qquad$S_1, S_2 \gets$ split$(S^*, c)$
    \State \qquad $Q_{k+1} \gets (Q_k \setminus \{S^*\}) \cup \{S_1, S_2\}$
    \State \qquad$k \gets k+1$
    \State \textbf{end while}
    \State \textbf{Output:} $x^*$
    \end{algorithmic}
\end{algorithm}

The algorithm begins by initializing a simplex \( S_0 \), which contains the feasible values of the sampling rate allocations \( \alpha_i \) across the workers. Each vertex \( V_i \) of the simplex corresponds to a specific allocation vector that satisfies the constraint \( \sum_{i=1}^n \alpha_i \leq C \) with \( \alpha_i \geq 0 \) for all \( i \). In our implementation, we construct the initial simplex \( S_0 \) with \( n+1 \) vertices in \( \mathbb{R}^n \), where each vertex corresponds to a deterministic allocation strategy. Specifically, the first vertex is the zero vector \( V_0 = \mathbf{0} \), representing no sampling for any worker. The remaining \( n \) vertices are defined as \( V_i = C \cdot \mathbf{e}_i \) for \( i = 1, \dots, n \), where \( \mathbf{e}_i \) is the \( i \)th standard basis vector in \( \mathbb{R}^n \). That is, each \( V_i \) allocates the entire sampling budget \( C \) to a single worker \( i \), while all other workers receive zero sampling rates. This construction ensures that the convex hull of the simplex covers the relevant feasible region of the optimization problem and provides a structured starting point for the algorithm. In the sum-of-ratios formulation given in (\ref{problem2}), we define for each worker $i$, functions $f_i(\alpha_i)$ and $g_i(\alpha_i)$ corresponding to the numerator and denominator of the objective function, respectively.

The global objective is then to maximize the total task success rate allocation across all workers, that is, $\sum_{i=1}^n \frac{f_i(\alpha_i)}{g_i(\alpha_i)}$ in (\ref{problem2}). At each iteration of the algorithm, the \textit{upper bound} \( UB(S) \) (which is in the form of a linear optimization problem) is computed by solving a relaxed version of the original sum-of-ratios problem over the current simplex \( S \), denoted as \( P_3(S) \). A \textit{lower bound} \(LB(S)\) is obtained by evaluating the global objective at the feasible point derived from the solution of the relaxed problem \(P_3(S)\). The algorithm continues to refine the simplex set until the stopping condition
\begin{align}
 UB(S) - LB(S) \leq \rho \cdot UB(S)   
\end{align}
is met, where \( \rho > 0 \) is an arbitrarily small predetermined tolerance level. If the stopping condition is not satisfied, the algorithm identifies the longest edge of the current simplex and bisects it to generate two new sub-simplexes. These sub-simplexes replace the original one in the active set, and the process repeats. This iterative partitioning continues until convergence, progressively narrowing down the feasible region and improving the quality of the global solution estimate.

For a given set of $p_i$'s, although the optimization problem in (\ref{problem2}) is non-convex in $\alpha_i$'s, with Algorithm~\ref{Alg1}, we can solve it globally optimally. Then, for a given set of $\alpha_i$'s, the optimization problem in (\ref{problem2}) in terms of $p_i$'s  has the same structure. For that, we again apply Algorithm~\ref{Alg1} to optimally solve for $p_i$'s. We repeat this process until $\alpha_i$'s and $p_i$'s converge. Although the overall iterative solution between $\alpha_i$'s and $p_i$'s is sub-optimal, we show in the next section that our proposed solution performs well compared to the global optimal solution attained by exhaustive search.       

\section{Numerical Results}\label{sect:numerical results}
In this section, we present numerical results to evaluate the task allocation strategies derived in Sections~\ref{sect:most} and \ref{sect:moderate}. 

\begin{figure}[t]
    \centering
    \includegraphics[width=0.67\linewidth]{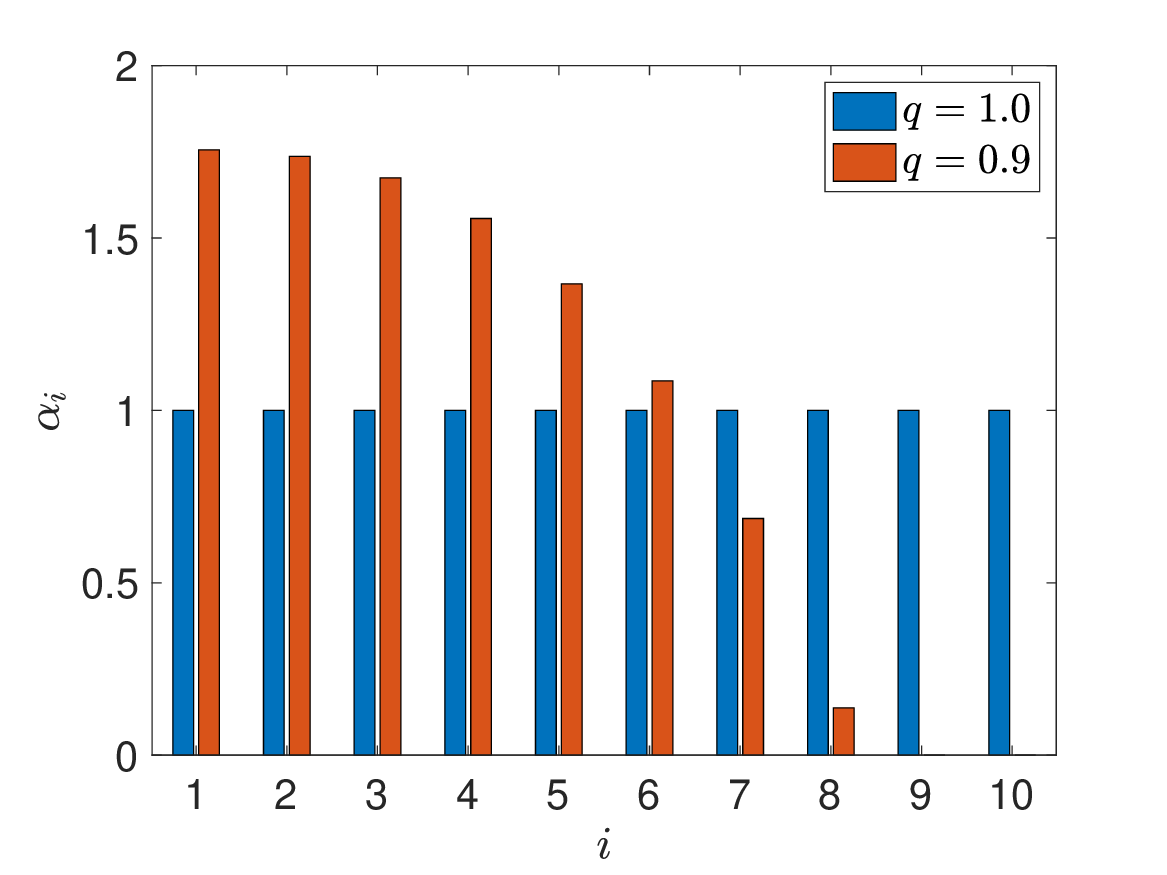}
    \vspace{-0.2cm}
    \caption{Optimal sampling rate allocations for $n = 10$, $C = 10$, and $\mu_i = 1$. The uniform and heterogeneous recovery rate configurations correspond to $q = 1.0$ and $q = 0.9$, respectively. In both cases, the recovery rates are normalized such that $\sum_i \lambda_i = 20$.}
    \label{fig:example1}
    \vspace{-0.5cm}
\end{figure}

In the first numerical simulation, we consider a system setting described in Section~\ref{sect:most} where the source assigns tasks to workers only when they are most efficient, that is, only when $s_i(t) =3$. In this example, the source assigns tasks to \( n = 10 \) workers, where each worker has an exhaustion rate of \( \mu_i = 1 \). The recovery rates are modeled as $\lambda_i = b q^{i-1}$ for all $i$, where we consider two different values for \( q \in \{1.0, 0.9\} \). We set the $b$ values such that in both cases, we have \( \sum_i \lambda_i = 20 \), allowing a fair comparison. The values of \( q \) that are close to 1 imply more homogeneous workers, while smaller values of \( q \) introduce variation in recovery rates across the worker population. The total sampling rate budget is constrained by \( C = 10 \). We compute the optimal sampling rates $\alpha_i$'s using the solution in~(\ref{eqn:alpha_i_final}) and evaluate the corresponding objective value defined in~(\ref{problem1}). The resulting sampling allocations for both settings are illustrated in Fig.~\ref{fig:example1}, and the associated performance metrics are summarized below.

For \( q = 1.0 \), the system achieves a utility of 1.7391, which decreases to 1.2967 when \( q = 0.9 \). This decline illustrates how task processing efficiency deteriorates as worker recovery rates become more heterogeneous. In the case of \( q = 1.0 \), all workers have identical recovery behavior, leading to a uniform distribution of the sampling budget and a greater aggregate task success. However, when \( q = 0.9 \), the recovery rates geometrically decrease with the worker index, and the optimal allocation policy avoids assigning any sampling budget to workers with poor recovery potential. This demonstrates the threshold-based behavior of the optimal solution, which prioritizes efficient workers while disregarding less favorable ones under the given resource constraint.

In the second numerical example, we consider a task allocation scenario with \( n = 2 \) workers, where the source allocates a total sampling budget of \( C = 10 \) across them. We focus on the setting in Section~\ref{sect:moderate}, where tasks can be assigned not only in fully efficient states but also in moderately efficient states as well, following the formulation in \eqref{problem2}. The recovery and exhaustion rates of the workers are chosen as \( \lambda = \{10, 20\} \) and \( \mu = \{5, 1\} \), respectively. We investigate how the probability of success of moderate-efficiency tasks \( p_s \in [0,1] \) influences the optimal task assignments and sampling rates. 

\begin{figure}[t]
    \begin{center}
        \subfigure[]{%
        \includegraphics[width=0.49\columnwidth]{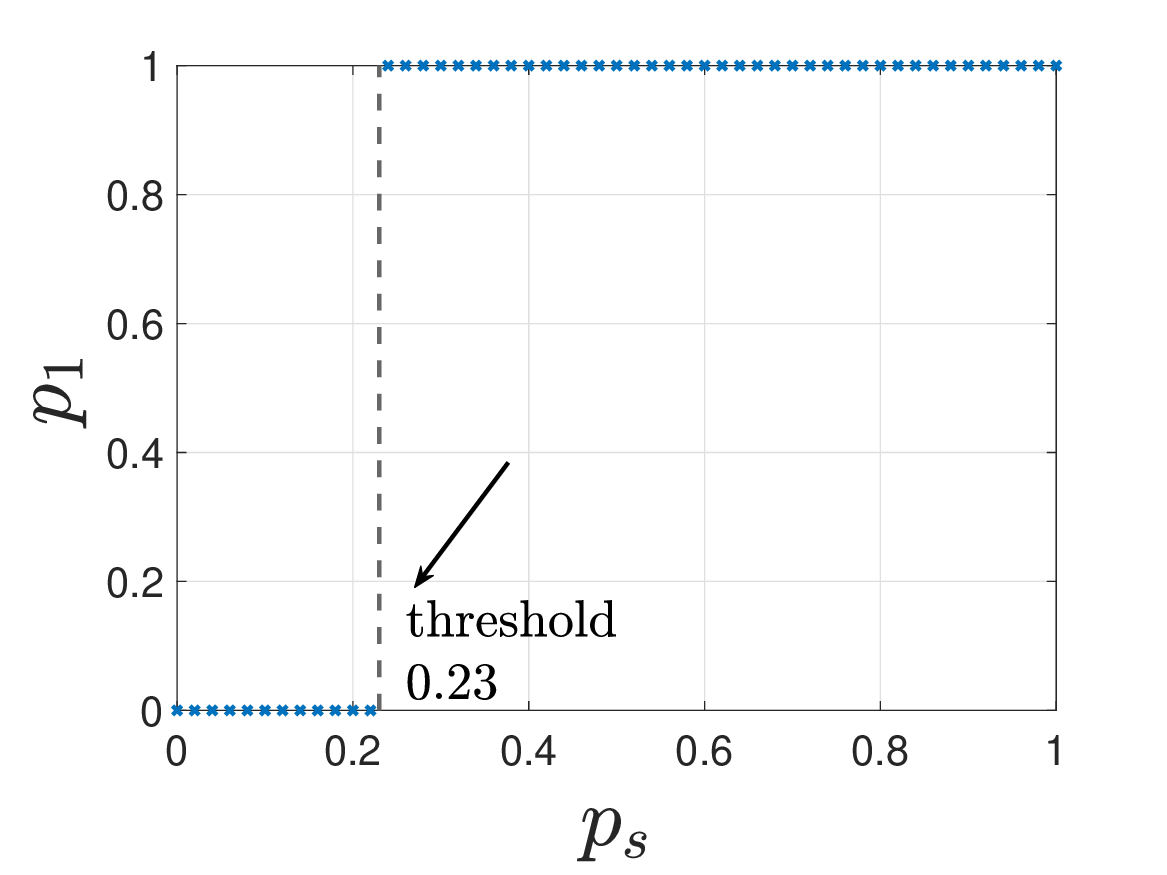}}
        \subfigure[]{%
        \includegraphics[width=0.49\columnwidth]{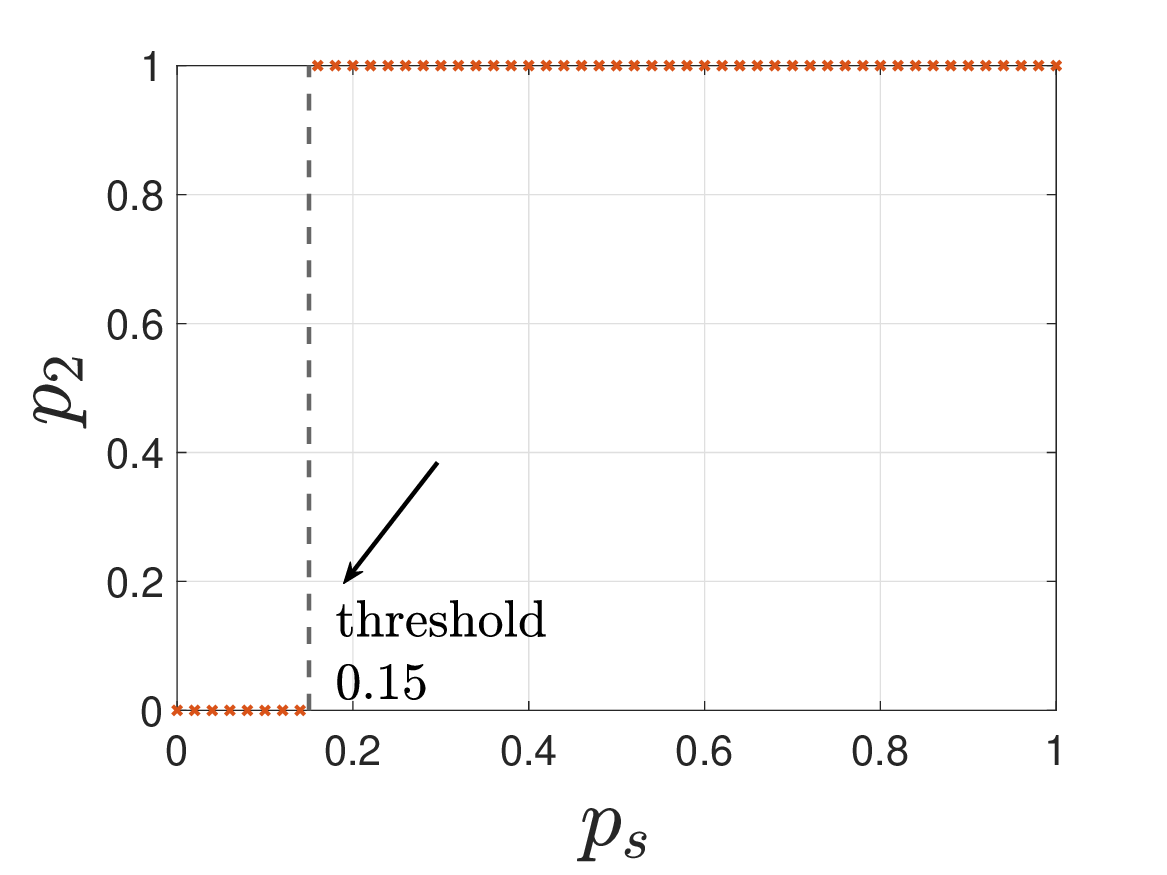}}
    \end{center}
    \vspace{-0.3cm}
    \caption{Optimal task assignment probabilities in the moderately efficient state as functions of \(p_s\), for a system with \(n = 2\) workers and total sampling budget \(C = 10\). Thresholds occur at \(p_s = 0.23\) and \(p_s = 0.15\), respectively.}
    \label{fig:example2}
\end{figure}

Algorithm~\ref{Alg1} is implemented to compute \( \alpha_1 \) and \( \alpha_2 \) for each combination of \( (p_s, p_1, p_2)\in [0,1]^3 \), and the parameters constituting the highest utility is obtained. The results of the simulation are illustrated in Fig.~\ref{fig:example2} in which both \( p_1 \) and \( p_2 \) exhibit a threshold behavior according to \( p_s \). The assignment probabilities \( p_i \) remain zero until a certain threshold is reached, beyond which they jump to one. This behavior is indicative of a policy switching strategy: below the threshold, it is best to assign tasks exclusively when the worker is in a fully efficient condition; above the threshold, it is advantageous to utilize moderately efficient state. We observe that the ratio of each worker's \( \lambda_i / \mu_i \) determines the threshold value at which this transition takes place. For example, the threshold for worker~1 occurs at \( p_s = 0.15 \), whereas it is \( p_s = 0.23 \) for worker~2. Consequently, the worker with the higher \( \lambda_i / \mu_i \) ratio (i.e., faster recovery) crosses the threshold at a lower \( p_s \), highlighting that it is more advantageous to assign tasks in moderate states. We also note that the threshold-based structure observed here simplifies implementation as the optimal policy for each worker $i$ reduces to a binary decision based on whether \( p_{s,i} \) exceeds the corresponding threshold or not.

\begin{figure}[t]
    \centering
    \includegraphics[width=0.7\columnwidth]{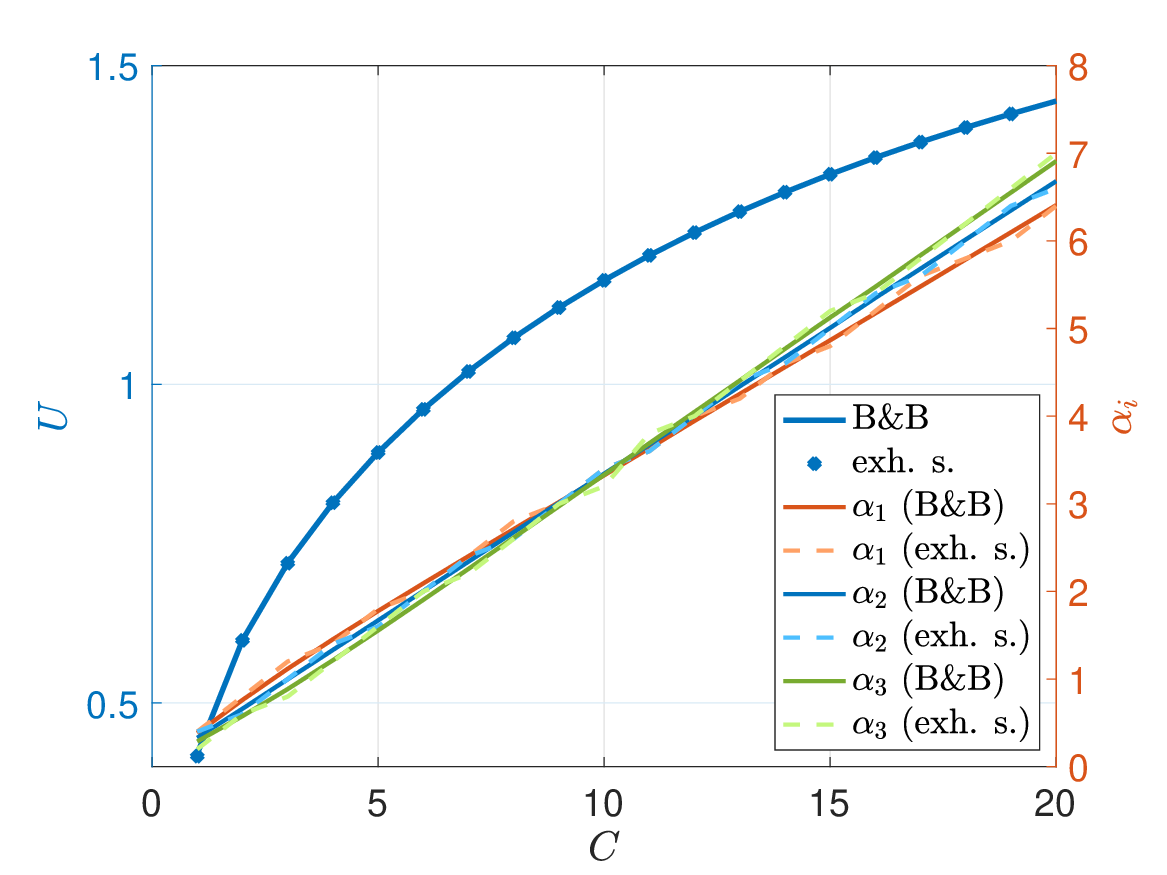}
    \vspace{-0.2cm}
    \caption{Comparison of the optimal utility values obtained by two methods: a branch-and-bound algorithm in Algorithm~\ref{Alg1} and an exhaustive search algorithm with the system of $n = 3$ workers.}
    \label{fig:utility_vs_budget}
    \vspace{-0.5cm}
\end{figure}

In the third simulation result, we validate the performance of the branch-and-bound algorithm stated in Algorithm~\ref{Alg1} by implementing an exhaustive search algorithm. We consider a system of $n = 3$ workers, with recovery rates $\lambda = \{2.5, 3.0, 3.5\}$, exhaustion rates $\mu = \{1, 1, 1\}$, and task success probabilities $p_s = \{0.7, 0.7, 0.7\}$. The total sampling budget $C$ is varied from $1$ to $20$. For each value of $C$, we first solve for the optimal sampling rates $\alpha_i$ under the assumption of fixed task assignment probabilities. Then, given these $\alpha_i$'s, we iteratively optimize the task assignment probabilities $p_i$'s, and refine the allocation to obtain the final optimal values of both $\alpha_i$'s and $p_i$'s. The resulting utilities are shown in Fig.~\ref{fig:utility_vs_budget} for both methods. The two approaches yield matching results across the entire range of $C$, validating the performance of the branch-and-bound algorithm. Moreover, we observe that the optimal $p_i$ values converge to $1$ for all workers, as in the previous example due to the fact that all workers exceed the threshold required for task assignment.

\section{Conclusion}
In this work, we considered the efficient task scheduling problem to workers that can exhaust themselves as a result of processing a task. Here, we modeled the workers to have three internal states: exhausted, moderately efficient, and the most efficient states. We studied two settings where the source can assign tasks to the workers i) only when they are most efficient, and ii) when they are also moderately efficient. For the first part, we showed that the optimal rate allocation policy is a threshold policy whereas for the second part, we provided an algorithm which iteratively finds an suboptimal allocation policy that performs well compared to the globally optimal solution. Future research can explore a threshold structure for the selection of $p_i$'s to improve the second part, and further explore MDP-based scheduling methods.       

\bibliographystyle{unsrt}
\bibliography{IEEEabrv, myLibrary}

\begin{thebibliography}{10}

\bibitem{Bochkovskyi2024workerfatigue}
A.~P. Bochkovskyi and N.~Y. Sapozhnikova.
\newblock Stochastic models of risk management of worker fatigue emergence.
\newblock {\em Journal of Achievements in Materials and Manufacturing Engineering}, 2024.

\bibitem{Kumar2023}
K.~Kumar, M.~Jain, and C.~Shekhar.
\newblock Machine repair system with threshold recovery policy, unreliable servers and phase repairs.
\newblock {\em Quality Technology and Quantitative Management}, July 2023.

\bibitem{cholette2014machine}
M.~E. Cholette and D.~Djurdjanovic.
\newblock Degradation modeling and monitoring of machines using operation-specific hidden {M}arkov models.
\newblock {\em IIE Trans.}, 46(10):1107--1123, 2014.

\bibitem{scheuvens2021control}
L.~Scheuvens, T.~Hößler, P.~Schulz, N.~Franchi, A.~N. Barreto, and G.~Fettweis.
\newblock State-aware resource allocation for wireless closed-loop control systems.
\newblock {\em IEEE Trans. Commun.}, 69(9):5622--5636, 2021.

\bibitem{channels_that_die}
L.~R. Varshney, S.~K. Mitter, and V.~K. Goyal.
\newblock Channels that die.
\newblock In {\em Allerton Conference}, September 2009.

\bibitem{zhou2022server}
J.~Zhou, K.~Cao, X.~Zhou, M.~Chen, T.~Wei, and S.~Hu.
\newblock Throughput-conscious energy allocation and reliability-aware task assignment for renewable powered in-situ server systems.
\newblock {\em IEEE Trans. Comput.-Aided Des. Integr. Circuits Syst.}, 41(3):516--529, 2022.

\bibitem{chen2014energyharvest}
J.~Chen, T.~Wei, and J.~Liang.
\newblock State-aware dynamic frequency selection scheme for energy-harvesting real-time systems.
\newblock {\em IEEE Trans. Very Large Scale Integr. (VLSI) Syst.}, 22:1679--1692, August 2014.

\bibitem{Yates2021JSAC}
R.~D. Yates, Y.~Sun, D.~R. Brown, S.~K. Kaul, E.~Modiano, and S.~Ulukus.
\newblock Age of information: An introduction and survey.
\newblock {\em IEEE Journal on Selected Areas in Comm.}, 39(5):1183--1210, May 2021.

\bibitem{Maatouk_remote_est}
A.~Maatouk, S.~Kriouile, M.~Assaad, and A.~Ephremides.
\newblock The age of incorrect information: A new performance metric for status updates.
\newblock {\em IEEE/ACM Transactions on Networking}, 28(5):2215--2228, 2020.

\bibitem{Akar2024_CTMC}
N.~Akar and S.~Ulukus.
\newblock Query-based sampling of heterogeneous {CTMCs}: Modeling and optimization with binary freshness.
\newblock {\em IEEE Transactions on Communications}, 72(12):7705--7714, 2024.

\bibitem{Pappas2024c}
J.~Luo and N.~Pappas.
\newblock Minimizing the age of missed and false alarms in remote estimation of {Markov} sources.
\newblock In {\em MobiHoc}, 2024.

\bibitem{Pappas2024b}
M.~Salimnejad, M.~Kountouris, A.~Ephremides, and N.~Pappas.
\newblock Version innovation age and age of incorrect version for monitoring {Markovian} sources.
\newblock In {\em WiOpt}, 2024.

\bibitem{Pappas2024a}
J.~Luo and N.~Pappas.
\newblock Semantic-aware remote estimation of multiple {Markov} sources under constraints.
\newblock In {\em WiOpt}, 2024.

\bibitem{liyanaarachchi2025mm}
S.~Liyanaarachchi and S.~Ulukus.
\newblock Optimum monitoring and job assignment with multiple {Markov} machines.
\newblock {\em arXiv:2501.18572}, 2025.

\bibitem{banerjee2025mm}
S.~Banerjee and S.~Ulukus.
\newblock Tracking and assigning jobs to a {M}arkov machine.
\newblock In {\em IEEE Infocom}, 2025.

\bibitem{Tang_MDP}
H.~Tang, J.~Wang, L.~Song, and J.~Song.
\newblock Minimizing age of information with power constraints: Multi-user opportunistic scheduling in multi-state time-varying channels.
\newblock {\em IEEE JSAC}, 38(5):854--868, 2020.

\bibitem{Boyd04}
S.~P. Boyd and L.~Vandenberghe.
\newblock {\em Convex Optimization}.
\newblock Cambridge University Press, 2004.

\bibitem{bastopcu_binary_freshness_21}
M.~Bastopcu and S.~Ulukus.
\newblock Information freshness in cache updating systems.
\newblock {\em IEEE Trans. on Wireless Comm.}, 20(3):1861--1874, 2021.

\bibitem{jong2012practical}
Y.~Jong.
\newblock Practical global optimization algorithm for the sum-of-ratios problem.
\newblock {\em arXiv preprint arXiv:1207.1153}, 2012.

\bibitem{khorramabadi2021sum}
S.~S. Khorramabadi.
\newblock Sum of ratios optimization using a new variant of dinkelbach's algorithm.
\newblock 2021.

\bibitem{SHEN20132301}
P.~Shen, W.~Li, and X.~Bai.
\newblock Maximizing for the sum of ratios of two convex functions over a convex set.
\newblock {\em Computers \& Operations Research}, 40(10):2301--2307, 2013.

\bibitem{SHEN2009145}
P.-P. Shen, Y.-P. Duan, and Y.-G. Pei.
\newblock A simplicial branch and duality bound algorithm for the sum of convex–convex ratios problem.
\newblock {\em Journal of Computational and Applied Mathematics}, 223(1):145--158, 2009.

\end{thebibliography}

\end{document}